\documentclass[runningheads]{llncs}
\usepackage{graphicx}
\begin{document}
\title{Short Paper: Privacy Comparison of Contact Tracing Mobile Applications for COVID-19}
%
%
\author{Mohamed-Lamine Messai\inst{1} \and
Hamida Seba\inst{2}} 
\authorrunning{M-L. Messai et al.}
%
\institute{Ecole d'Ingénieur CPE Lyon, Université de Lyon, France. \\
\email{mohamed-lamine.messai@cpe.fr}\\
\and  Université de Lyon, CNRS, Université Lyon 1
LIRIS, UMR5205, France.  \\
\email{hamida.seba@univ-lyon1.fr}}
\maketitle              
\begin{abstract}
With the COVID-19 pandemic, quarantines took place across the globe. In the aim of stopping or slowing the progression of the COVID-19 contamination, many countries have deployed a contact tracing system to notify persons that be in contact with a COVID-positive person. The contact tracing system is implemented in a mobile application and leverages technologies such as Bluetooth to trace interactions between persons. This paper discusses different smart-phone applications based on contact tracing system from privacy point of view.

\end{abstract}
\section{Introduction}
The globe is sick by the COVID-19 epidemic. Many countries are choosing numeric solutions to help decreasing the virus propagation. They are implementing and deploying mobile applications with a system for tracking contacts of persons. Proposed applications leverage Bluetooth technology to trace interactions between persons.
The tracing system help the health authorities in these countries to decrease the number of contaminated persons. If a person is tested COVID-positive, it will be possible with such an application to alert all the persons that have been in physical or close contact with the infected person the previous days. 

Contact tracing is the process of identifying and monitoring which persons had contact with infectious persons. This is possible by leveraging our smart-phones. The smart-phone of a user broadcasts its identifier which is recorded by nearby smart-phones of crossed persons. If a user tests positive for the COVID-19, the list of recorded identifiers by that user is used to notify crossed persons that they were in close range from an infected person. The notified persons are invited to do varying courses of actions including self-quarantine, symptom watch and testing \cite{key4}. 

Contact tracing systems use Bluetooth Low Energy (BLE) technology which was first specified in 2010. The BLE has become the dominant low-power wireless technology in a variety of consumer
electronics devices, including the smart-phones. In fact, smart-phones bring widespread presence to BLE. The energy-efficiency of the BLE allows its use in contact tracing systems. To note that, other smart-phone application solutions for COVID-19 pandemic based on the use of GPS are developed such as Mapy.cz in Czech and the contact tracing application from Iran \cite{key6}. It is proposed to the users to archive the GPS data of their movements and to share it with the health investigators if necessary \cite{key5}. The use of GPS makes the application a battery-intensive solution and it does not work well indoors.

In this paper, we discuss the privacy issues with contact-tracing apps-- we set out for a privacy comparison of the most relevant contact tracing applications. We first discuss the privacy consideration in Section 2. Then, we describe briefly each of the contact tracing application in Section 3 and compare them according to their privacy property in Section 4. In Section 5, we conclude the paper.
\section{Privacy Consideration}
The privacy is a state in which one is not observed or disturbed by other persons. A contact tracing application is a mobile phone application which records, using Bluetooth, contacts between persons, in order to detect a possible risk of infection. We remark that there is a risk for privacy violation which can be mitigated more or less
well, but not completely, depending on the approach taken in the application design such as anonymization technique \cite{key1,key7}.

\subsection{Privacy threat sources}
In a country, the health authority is assumed to offer the medical tests if a person is
infected or not. The health authority, without respecting the privacy of citizens, can try to deanonymize users' IDs to trace possible new infections. It can also try to use collected data to identify contacts of persons in other kind of applications. Another source of threat is a malicious application developer which could add a snooping module on data from other applications on the smart-phone or requesting permissions which may result in surveillance of the user. Other kinds of malicious actions might be supposed. Infected users are also a source of threat. It can try to spread panic by wandering around to distort the application or set IDs as past contacts, even though they have not met. This could lead into error the corresponding users. The last source of threat is the hacker who accesses the user's smart-phone or the server that store the collected contact tracing data. The attacker can analyze these data or sell them to third parties.




\section{Contact tracing applications}
In the aim to break infection chains, several mobile applications are developed supporting contact tracing. In this section, the most relevant applications are described. 

Singapore was one of the first nations to adopt a Bluetooth-powered contact-tracing application, fueling plenty of global debate about the best way to deploy the technology. The application is called TraceTogether \cite{key3}. TraceTogether is an application that can be downloaded voluntarily and facilitates the contact tracing process. With your consent, it exchanges Bluetooth signals with nearby smart-phones running the same application. When a person sign up on the TraceTogether, a random User ID is generated and associated with your mobile number. Both the mobile number and User ID are stored in a secure server, and never shown to the public \cite{key3}. AC19 is the Iranian application. It requires a user to register using their mobile number. It uses unencrypted GPS data for contact tracing. While Asian countries had notable success using mobile applications for tracing infections, these applications are highly invasive with respect to the user's privacy \cite{key6}. For instance, India is the only democracy making its application called Aarogya Setu mandatory for millions of people. As a consequence, a considerable effort must be done to present privacy preserving tracing applications. In Europe, several variants of applications are proposed based on the PEPP-PT (Pan-European Privacy-Preserving Proximity Tracing) framework. The French application is called StopCovid. The German one is NTK as described in \cite{key2}. They are closely related and thus have very similar privacy properties. In Austria, Stopp Corona is the proposed application and it is also based on the PEPP-PT framework. Another interesting collaboration effort is the Private Automated Contact Tracing (PACT), led by MIT researchers. In USA, different contact tracing applications have emerged recently. One of these application is Covid-Watch based on the use of Bluetooth. When a user is infected then the list of anonymous contact numbers in his smart-phone is publicly made available with her consent. Other users
then use this public list to find out whether they came in close contact of an infected person or not. Decentralized Privacy-Preserving Proximity Tracing (DP-3T) is an open source application listed as one of several privacy-preserving decentralized approaches to contact tracing. Other countries are also developing their own contact tracing application. Generally, the application uses GPS based location tracking, Bluetooth or a combination of them for collecting information.

\section{Comparison}
The following table presents a comparison of contact tracing applications described previously regarding the sources of privacy threat defined in Section 2. A unique identification number is randomly generated in the majority of discussed applications to ensure anonymity and it is used for tracing all the contacts in close proximity of users. An unavoidable privacy loss of all application in the scenario where a user has only contact with one other user, then the user can learn whether the other person is infected based on whether she/he receives a COVID-19 notification. 
\begin{table*}[]

\tiny
\caption{Tracing contact applications comparison.}
\begin{tabular}{|c|c|c|c|c|c|c|}
\hline
Apps          & Used technology & Need trust server & False claims for users & Anonymization &                           Malicious developer & Data destruction\\ \hline
Aarogya Setu  & BLE and GPS      & Yes               & Yes                    & No            & Not open source & Yes    \\ \hline
TraceTogether & BLE              & Yes                & Yes                    & Yes           & Open source  & Yes       \\ \hline
Covid-Watch   & BLE              & No                & Yes                    & Yes           & Not open source  &  N.C.  \\ \hline
Stopp Corona  & BLE and GPS      & No               & Yes                    & Yes           & Open source  & Yes       \\ \hline
AC19          & GPS              & Yes               & Yes                    & No            & Not open source  & N.C.   \\ \hline
PACT          & BLE              & No                & Yes                    & Yes           & Not open source  & N.C.   \\ \hline
StopCovid        & BLE              & Yes               & Yes                    & Yes           & Open source   & Yes  \\ \hline
\end{tabular}

\end{table*}

No matter how smart-phone contact tracing applications might be helpful to break infection chains, it still come with privacy concerns. Whether the smart-phone application is base on GPS or Bluetooth, it is plenty accurate to invade privacy and reveal your personal information  \cite{key2}. Applications that need a trust server are the applications based on the centralized approach where the server determines who has been exposed to COVID-19. The drawback of a centralized approach is that the server is the single point of failure. Applications that do not need a trust server are the applications based on the decentralized approach. Data destruction determines even collected personal data are deleted after a period of time (N.C. for Not Communicated).
  
\section{Conclusion and future work}
Across the globe, scientists and researchers of the fields of information technology are working to develop contact tracing applications to outbreak COVID-19 pandemic. A special care must be taken for security and privacy. Furthermore, some scientists urge that the health benefits of deploying mobile applications with system tracing need to be analyzed. In this poster, we present a preliminary comparison study of privacy-preserving contact tracing mobile applications for COVID-19.
Contact tracking is a very central topic in this moment. As a future work, we plane to analyse the privacy issue in terms of data acquisition, transmission and storage.
%
%
%
 \bibliographystyle{splncs04}
\bibliography{mybibliography}

\begin{thebibliography}{1}
\providecommand{\url}[1]{\texttt{#1}}
\providecommand{\urlprefix}{URL }
\providecommand{\doi}[1]{https://doi.org/#1}

\bibitem{key2}
Aisec, F.: Pandemic contact tracing apps: Dp-3t, pepp-pt ntk, and robert from a
  privacy perspective. Cryptology ePrint Archive, Report 2020/489 (2020),
  \url{https://eprint.iacr.org/2020/489}

\bibitem{key5}
Alwashmi, M.F.: The use of digital health in the detection and management of
  covid-19. International Journal of Environmental Research and Public Health
  \textbf{17}(8), ~2906 (2020)

\bibitem{key3}
Cho, H., Ippolito, D., Yu, Y.W.: Contact tracing mobile apps for covid-19:
  Privacy considerations and related trade-offs. arXiv preprint
  arXiv:2003.11511  (2020)

\bibitem{key4}
Oliver, N., Lepri, B., Sterly, H., Lambiotte, R., Delataille, S., De~Nadai, M.,
  Letouz{\'e}, E., Salah, A.A., Benjamins, R., Cattuto, C., et~al.: Mobile
  phone data for informing public health actions across the covid-19 pandemic
  life cycle (2020)

\bibitem{key6}
Shukla, M., Lodha, S., Shroff, G., Raskar, R., et~al.: Privacy guidelines for
  contact tracing applications. arXiv preprint arXiv:2004.13328  (2020)

\bibitem{key1}
Tang, Q.: Privacy-preserving contact tracing: current solutions and open
  questions. arXiv preprint arXiv:2004.06818  (2020)

\bibitem{key7}
Trieu, N., Shehata, K., Saxena, P., Shokri, R., Song, D.: Epione: Lightweight
  contact tracing with strong privacy. arXiv preprint arXiv:2004.13293  (2020)

\end{thebibliography}

\end{document}